\documentclass{jpp}
\usepackage{graphicx}
\usepackage{epstopdf, epsfig}

\newtheorem{mydef}{Definition}

\shorttitle{Lagrangian coherent structures and plasma transport processes}
\shortauthor{M. V. Falessi, F. Pegoraro and T.J. Schep}

\title{Lagrangian coherent structures and plasma transport processes}

\author{M. V. Falessi\aff{1}
  \corresp{\email{falessi@fis.uniroma3}},
  F. Pegoraro\aff{2}
 \and T. J.  Schep\aff{3}}

\affiliation{\aff{1} Dipartimento di Matematica e Fisica, Roma Tre University, Via della Vasca Navale 84, 00199 Roma, Italy
\aff{2}Dipartimento di Fisica, Pisa University, Largo Bruno Pontecorvo 3, 56127 Pisa, Italy
\aff{3}Department of Physics, Eindhoven University of Technology, De Rondom 70,5612 AP Eindhoven, Netherlands}

\begin{document}

\maketitle

\begin{abstract}
A dynamical system framework is  used to  describe  transport processes in plasmas embedded in a magnetic field.
For periodic systems with one degree of freedom the Poincar\'e map  provides a splitting of the phase space into regions where particles have different kinds of motion: periodic, quasi-periodic or chaotic. The boundaries of these regions are transport barriers; i.e., a trajectory cannot cross such boundaries during the whole evolution of the system.  Lagrangian Coherent Structure (LCS)  generalize this method to systems with the most general time dependence, splitting the phase space into regions with different qualitative behaviours. This leads to the definition of finite-time transport barriers, i.e. trajectories cannot cross the barrier for a finite amount of time. This methodology can be used to identify fast recirculating regions in the dynamical system and to characterize  the transport between them.
\end{abstract}

\section{Introduction}
Transport  phenomena are ubiquitous in nature and involve the redistribution of physical quantities such as mass, charge, linear and  angular momentum and energy etc. Different mechanisms are at work in the transport processes ranging from diffusion to advection and mixing in the case of turbulent or chaotic motions. Recently a new concept in the study of transport processes in complex fluid flows was  introduced by \citet{peacock2013lagrangian}:  Lagrangian coherent structures (LCS). In a two-dimensional configuration these structures are special lines\footnote{In the most general system the LCS are surfaces advected by the flow but here for the sake of simplicity we will only deal with $2D$ systems.} advected by the fluid which organize the flow transport processes by attracting or repelling the nearby fluid elements over a finite time span. Strictly speaking we should call these structures Hyperbolic LCS. Two other kinds of LCS have been introduced in the literature. For the definition of all the different kinds of LCS we indicate the recent review by \citet{haller2015lagrangian}. For the sake of simplicity we will refer to these structures simply as LCS. These special lines organise the flow splitting the domain into macro-regions with fast mixing phenomena inside them. Over the finite time span which characterizes the LCS these macro regions do not exchange fluid elements. The LCS have been widely used in the literature to characterize transport processes in various systems: the pollutant transport on the ocean surface \citep{coulliette2007optimal}, blood flow \citep{shadden2008characterization},  the spreading of
plankton blooms \citep{huhn2012impact}, turbulent combustion \citep{hamlington2011interactions}, jellyfish predator-prey interaction \citep{peng2009transport}, atmospheric dataset analysis \citep{tang2010lagrangian}, solar photospheric flows \citep{chian2014detection}, saturation of a nonlinear dynamo \citep{rempel2013coherent}, etc.

Plasmas are often studied using ``fluid''  theories either in phase space, such as the Vlasov-Maxwell system, or in physical space such  as  the  two fluid  and the MHD systems. The LCS techniques can therefore  be  applied to study transport processes, i.e. the mixing of fluid elements, in these systems. In  (citation to another proceeding: Carlevaro, Falessi, Montani, Zonca) the LCS has been used to quantify the phase space transport due to the interactions between two supra-thermal electron beams and a cold, homogeneous, background plasma. In two recent works \citep{borgogno2011barriersa, borgogno2011barriers} it was shown how the LCS can provide information about the electron transport due to the stochastization  of the magnetic field in a collisionless reconnection process. 

The introduction of these techniques is  relatively recent and, in spite of their increasing  use, their rigorous definition has been subject to debate \citep{shadden, haller2011variational}. The first definitions of an LCS were based on the Finite time Lyapunov exponent profile. These have been shown to be incorrect by G.Haller who found several counterexamples to this heuristic definition. The rigorous definition of an LCS as the most repulsive or attractive material line with respect to the nearby ones was introduced by \citep{haller2011variational}. In this article we provide a simplified version of this derivation in order to give to the reader some physical intuition about these structures, we analytically calculate the shape of the LCS in a simple Hamiltonian system and, finally, we compare the LCS obtained by \citet{borgogno2011barriersa} using the FTLE method with the ones obtained with the rigorous definition introduced. The numerical framework used to compute the LCS shape was  developed by \citet{onu2015lcs}.

\section{Lagrangian Coherent Structures and Transport Barriers}

\subsection{Lagrangian Coherent Structures as  most repelling material lines} 

Following \citep{haller2011variational},  in this section we  briefly review some mathematical concepts  that lead to  the definition of LCS. 

We consider a dynamical system  in 2D phase space ${\bf x} = (x,y)$,
\begin{equation}\label{1=}
\frac{d{x}}{dt}={v_x}(t,{x,y}),\qquad \frac{d{y}}{dt}={v_y}(t,{x,y})
\end{equation}
with  continuous differentiable flow map 
\begin{equation}\label{2=}
 {\boldsymbol \phi}_{t_0}^t({\bf {\boldsymbol  x}_0})={\bf x}(t,t_0,{\bf {\boldsymbol  x}_0}).
\end{equation}
Two neighbouring points ${\bf x}_0$ and ${\bf x}_0+\delta{\bf x}_0$ evolve into the points ${\bf x}$ and ${\bf x}+\delta{\bf x}$ under the linearized map
\begin{equation}\label{3=}
 |\delta x \rangle ={\boldsymbol   \nabla} {\boldsymbol \phi} _{t_0}^t \,  |\delta {  x}_0 \rangle .
\end{equation}
where for notational convenience, we adopt a  \emph{bra-ket} notation for vectors and scalar products and represent a generic column vector as $|c\rangle$ and a row vector as $\langle r |$. Their scalar product is denoted as $d=\langle c|r\rangle$.

Consider a curve $\gamma_0=\{{\bf x}_0=r(s)\}$. 
At  each point  ${\boldsymbol  x}_0 \in \gamma_0$,  define the unit tangent vector $|e_0 \rangle$ and the normal vector $|n_0\rangle$. In the  time interval $[t_0,t]$ the dynamics of the system advects the ``material line'' $\gamma_0$ into $\gamma_t$ and ${\boldsymbol  x}_0\in \gamma_0$ into ${\boldsymbol  x}_t \in \gamma_t$ . The linearized dynamics maps the tangent vector $\left|e_0\right \rangle$ into $\left|e_t\right \rangle$ which is tangent to $\gamma_t$ 
\begin{equation}
|e_t\rangle=\frac{\boldsymbol  {\nabla} {\boldsymbol \phi}^t_{t_0}({\boldsymbol  x}_0)|e_0\rangle}{ \sqrt{\left\langle e_0\left|\left(\boldsymbol  {\nabla} {\boldsymbol \phi}^t_{t_0}\right)^T\boldsymbol  {\nabla} {\boldsymbol \phi}^t_{t_0} \right|e_0\right\rangle}}\equiv \frac{\boldsymbol  {\nabla} {\boldsymbol \phi}^t_{t_0}({\boldsymbol  x}_0)|e_0\rangle}{ \sqrt{\left\langle e_0\left|C^t_{t_0}({\boldsymbol  x}_0)\right|e_0\right\rangle}},
\label{tangent}
\end{equation}
where $C^t_{t_0}({\boldsymbol  x}_0)\equiv \left(\boldsymbol  {\nabla} {\boldsymbol \phi}^t_{t_0}\right)^T \, \boldsymbol  {\nabla} {\boldsymbol \phi}^t_{t_0}$  is the \emph{Cauchy-Green strain tensor} and $^T$ stands for transposed. This symmetric tensor describes the deformation of an arbitrarily small circle of initial conditions, centered in ${\boldsymbol  x}_0$ caused by the  flow  in a time interval $[t_0,t]$. As an example let us  consider a circle centered in ${\boldsymbol  x}_0$ with radius $\|\delta {\boldsymbol  x}_0\|$. After the time interval $[t_0,t]$ this will be deformed into an ellipse with major axis in the direction of $\left|\xi_{max} \right\rangle$ and minor axis in the direction of $\left |\xi_{min} \right \rangle$ being $\left |\xi_{max}\right \rangle$ and $\left| \xi_{min}\right\rangle$ the two eigenvectors of $C^{t}_{t_0}({\boldsymbol  x}_0)$. The corresponding real and positive eigenvalues are $\lambda_{max}$ and $\lambda_{min}$. The length of the major axis is $\lambda_{max} \| {\delta } {\boldsymbol  x}_0 \| $ while the the length of the minor is $\lambda_{min}\|\delta {\boldsymbol  x}_0\|$. 
The curves with tangent vector along  $\left |\xi_{min} \right \rangle$  and, respectively,  $\left |\xi_{max} \right \rangle$  are called \emph{strain lines} of the Cauchy-Green tensor.
In general, the mapping  will not preserve the angle between vectors and therefore  usually $\left|n_t\right \rangle$  differs from $\boldsymbol  {\nabla} {\boldsymbol \phi}^t_{t_0} \left|n_0\right\rangle$. 

\begin{figure}
  \centerline{\includegraphics[width=0.7\columnwidth]{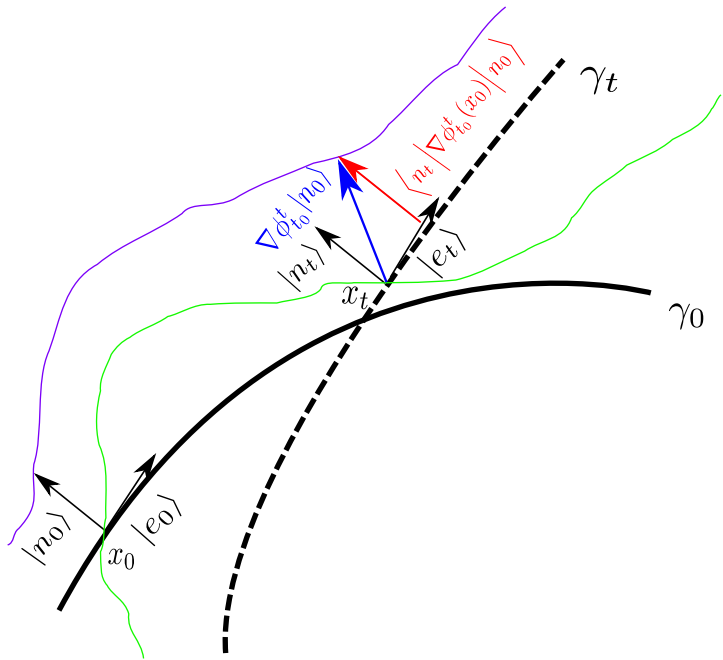}}
\caption{The green line and the purple  line represent the trajectories of two  elements  of the system. The continuous thick black line is  the curve $\gamma_0$ while the dashed one is its evolution  $\gamma_t$ at time $t$. We have marked in blue the evolution  of $\left | n_0 \right\rangle$ through the linearized dynamic and in red its projection on $\left | n_t \right\rangle$.}
\label{fig: a}
\end{figure}

Using the condition  \unexpanded{$ \left \langle n_0 | e_0 \right \rangle =\left \langle n_0\left|\boldsymbol  {\nabla}{\boldsymbol \phi}^{t_0}_t\boldsymbol  {\nabla}{\boldsymbol \phi}^{t}_{t_0}\right|e_0\right\rangle$=0} and inserting  Eq. (\ref{tangent}) we obtain the expression for $\left|n_t\right \rangle$ which is given by 
\begin{equation}
 |n_t\rangle=\frac{\left(\boldsymbol  {\nabla} {\boldsymbol \phi}^{t_0}_{t}\right)^T|n_0\rangle}{\sqrt{ \left\langle n_0\left|\boldsymbol  {\nabla} {\boldsymbol \phi}^{t_0}_{t}\left(\boldsymbol  {\nabla} {\boldsymbol \phi}^{t_0}_{t}\right)^T \right|n_0\right\rangle}}\equiv \frac{\left(\boldsymbol  {\nabla} {\boldsymbol \phi}^{t_0}_{t}\right)^T|n_0\rangle}{\sqrt{ \left\langle n_0\left|C^{-1} ({\boldsymbol  x}_0 )\right|n_0\right\rangle}},
\end{equation}
where $C^{-1} ({\boldsymbol  x}_0 ) = C^{t_0}_t ({\boldsymbol  x}_0 ) $ and the time interval marks have been  suppressed  as will be the case in the following formulae when not explicitly needed.
We  define the \emph{repulsion ratio} $\rho^t_{t_0}({\boldsymbol  x}_0,n_0)$ as the ratio at which material points, in other words points advected by the flow, initially taken  near the point ${\boldsymbol  x}_0 \in \gamma_0$,   increase  their distance from the curve in the time interval $[t_o,t]$: 
\begin{equation}
\rho^t_{t_0}({\boldsymbol  x}_0,n_0)=\left \langle n_t \left |\boldsymbol  {\nabla} {\boldsymbol \phi}^t_{t_0}({\boldsymbol  x}_0)\right| n_0\right \rangle.
\label{repulsionratio}
\end{equation}
In order to understand this definition we may imagine measuring the distance between ${\boldsymbol  x}_0$, taken on the curve $\gamma_0$, and a point initially placed at unit distance from the curve. After a sufficiently small amount of time  this distance is measured by the projection of the vector $\boldsymbol  {\nabla} {\boldsymbol \phi}^{t}_{t_0}({\boldsymbol  x}_0)|n_0\rangle$ along $\left| n_t \right \rangle$ as shown in Figure \ref{fig: a}.
Using the previous definitions, $\rho^t_{t_0}({\boldsymbol  x}_0,n_0)$ can be expressed either in terms of $n_0$ or of  $n_t$ as 
\begin{equation}
\rho^t_{t_0}({\boldsymbol  x}_0,n_0)=\frac{1}{\sqrt{\left \langle n_0 \left|C^{-1} ({\boldsymbol  x}_0)  \right | n_0 \right \rangle}}=\sqrt{\left \langle n_t \left |C ({\boldsymbol  x}_0) \right| n_t \right \rangle}.
\end{equation}
Similarly, the \emph{contraction rate} $L^{t}_{t_0}({\boldsymbol  x}_0)$ is proportional to the growth in time of the vector tangent to the material line
\begin{equation}
L({\boldsymbol  x}_0,e_0)= \sqrt{\left \langle e_0 \left| C ({\boldsymbol  x}_0) \right| e_0 \right \rangle} .
\end{equation} 
The aim in these definitions is to  characterize a LCS over a finite time interval $\left[t_0, t_0+T \right]$ as a material line, in other words a curve advected by the flow,  which is locally the strongest repelling 
or attracting curve with respect to the nearby ones. This leads, as shown by \citet{haller2011variational}, to the following definitions.
\begin{mydef}
A material line satisfying the following conditions at each point:
\begin{enumerate}   
\item \begin{equation} \lambda_{min}<\lambda_{max},\quad \lambda_{max}>1 \label{condizionelambda}\end{equation}
\item the tangent vector is along the eigenvector associated with the smallest eigenvalue
\begin{equation}
\left| e_0\right \rangle= \left |\xi_{min}\right \rangle, 
\label{condizione2}
\end{equation}
\item the gradient of the largest eigenvalue is along the curve
\begin{equation}
\left \langle \xi_{max}\Big| \nabla \lambda_{max}\right \rangle=0
\label{numericallyaproblem}
\end{equation}
\end{enumerate}
is called a Weak Lagrangian coherent structure (WLCS).
\end{mydef}

\begin{mydef}
\label{mostrepellingm}A WLCS which satisfies the following additional condition
\begin{enumerate}
\item at each point the  relationship 
\begin{equation}\label{maxrep}
\left \langle \xi_{max}\left |\nabla ^2 \lambda_{max}\right|\xi_{max}\right \rangle <0
\end{equation}
\end{enumerate}
holds is called a Lagrangian coherent structure.
\end{mydef}
A  simplified illustration  of the previous conditions can be given as follows.
The first condition is obtained when requiring that at  each point ${\boldsymbol  x}_0 $ of the material line the repulsion rate $\rho({\boldsymbol  x}_0,n_0)$ is larger than the contraction rate $L({\boldsymbol  x}_0,n_0)$ which represents the effect of the shear along the material line. At each point along  a  material line the tangent vector can be expressed in terms of the eigenvectors $ \left | \xi_{min}\right \rangle $ and $ \left | \xi_{max}\right \rangle $ of the \emph{Cauchy-Green} strain tensor $C$: 
\begin{equation}
\left |e_0\right \rangle=\alpha \left | \xi_{min}\right \rangle +\beta \left | \xi_{max} \right \rangle, \quad \alpha^2 +\beta^2=1.
\end{equation}
where $\alpha$ and $\beta$ represent the orientation of the material line and may be arbitrary functions of ${\boldsymbol  x}_0$. It follows that 
\begin{equation}
\left |n_0 \right \rangle= \alpha \left |\xi_{max}\right \rangle - \beta \left |\xi_{min}\right \rangle.
\end{equation}
We can therefore express the repulsion rate in terms of $\alpha$ and $\beta$ 
\begin{equation}
\rho^t_{t_0}({\boldsymbol  x}_0,n_0)=\frac{1}{\sqrt{\left \langle n_0 \left|C^{-1} \right | n_0 \right \rangle}}=\left(\frac{\alpha^2}{\lambda_{max}}+\frac{\beta^2}{\lambda_{min}}\right)^{-1/2}.
\end{equation}
Now  at each point  ${\boldsymbol  x}_0$ we maximize $\rho^t_{t_0}$ with respect to the direction 
of the tangent vector $\left |e_0 \right \rangle$ i.e., of the orientation of the chosen material line, 
with the constraint $\alpha^2 + \beta^2=1$. This yields  $\alpha=1$, $\beta=0$, so that
\begin{equation}
\rho^t_{t_0}=\sqrt{\lambda_{max}},
\label{repulsioneq}
\end{equation}  
which implies that the material line must be chosen to be the   strain line  oriented along the eigenvector $\left|\xi_{min} \right \rangle$ as required by  Eq. (\ref{condizione2}). 
At this stage we need  to maximize the repulsion rate along the material line with respect to nearby material lines. Therefore we define 
\begin{equation}
R^{t}_{t_0}(\gamma_0)\equiv \int_{s_1}^{s_2}ds \, \rho\left({\boldsymbol  x}_0(s),n_0(s)\right),
\label{functional}
\end{equation}
 which has the physical meaning of the repulsion rate integrated over a curve $\gamma_0$. We consider a curve $\gamma_{\epsilon}$ with points ${\boldsymbol x}_{\epsilon} \in \gamma_{\epsilon}$ such that 
\begin{equation}
{\boldsymbol x}_ \epsilon = {\boldsymbol  x}_0 +\epsilon h(s)\left|n_0\right \rangle.
\label{variedcurve}
\end{equation}
The first variation of Eq. (\ref{functional}) with respect to Eq. (\ref{variedcurve}) gives
\begin{equation}
  \delta R^{t}_{t_0}(\gamma_{\epsilon})[h]=\delta \int_{s_1}^{s_2}ds \, \rho\left(x_{\epsilon}(s),n_{\epsilon}(s)\right)=0,
\end{equation}
which can be computed as 
\begin{equation}
\lim_{\epsilon \to 0}\frac{\partial}{\partial \epsilon}\int_{s_1}^{s_2}ds\rho\left(x_{\epsilon}(s),n_{\epsilon}(s)\right)=\lim_{\epsilon \to 0}\frac{\partial}{\partial \epsilon}\int_{s_1}^{s_2}ds \sqrt{\lambda_{max}\left(x_{\epsilon}(s),n_{\epsilon}(s)\right)} \end{equation}
leading to 
\begin{equation}
\delta R^{t}_{t_0}(\gamma_{\epsilon})[h]=\int_{s_1}^{s_2}\frac{ds \, h(s)}{2\sqrt{\lambda_{max}}}\left \langle \left. \boldsymbol  {\nabla}\lambda_{max}\right | \xi_{max} \right \rangle = 0
\end{equation}
which vanishes if $\left |\boldsymbol  {\nabla}\lambda_{max} \right \rangle$ is  tangent to the material line. Requiring that this material line represents a maximum of the integrated repulsion rate we obtain 
\begin{equation}
\left \langle \xi_{max}\left|\nabla^2 \lambda_{max}\right|\xi_{max}\right \rangle<0
\end{equation}
which is the condition that defines  the locally most repelling LCSs. These structures are Lagrangian by definition and have no transport through them because they are material lines. 

\subsection{Lagrangian Coherent Structures as second derivative ridges} 

Second derivative ridges were defined by \citet{shadden} in terms of the features of the Lyapunov exponent field $\sigma ( {\boldsymbol  x}_0, t_0, t )$ that 
characterizes the rate of separation of close trajectories.

 The  finite time  Lyapunov exponent can be expressed  in terms of the Cauchy-Green tensor  eigenvalues as 
 \begin{equation}
 \label{Lyap}
 \sigma ( {\boldsymbol  x}_0, t_0, t ) = \frac{1}{ 2|t -t_0|}\, \ln{\lambda_{max}} ( {\boldsymbol  x}_0, t_0, t ) 
 \end{equation} 
 
 \begin{mydef}
Curves  ${\boldsymbol r}(s)$ (not necessarily   material lines) such that
\begin{enumerate} 
\item  the tangent vector  $\left| r^\prime(s) \right \rangle $and  $\,  {\boldsymbol \nabla}\sigma$ along the curve are parallel, 
\item the normal unit vector $\left| n \right \rangle$ is such that  along the curve  for all $ \left \langle u  |  u \right \rangle = 1$ 
 \begin{equation}
\left \langle  n  \left|  \Sigma \right| n \right \rangle = {\rm min}\, |_{ \langle u |  u \rangle = 1}  \label{Sigma}
\left \langle u  \left|  \Sigma \right| u \right \rangle <0 \end{equation}  
  \end{enumerate}
where $\Sigma $ is the Hessian matrix of the second derivatives of $\sigma$ with respect to $ {\boldsymbol  x}_0$
is called a second derivative ridge.
\end{mydef}
A major difference  between the two sets of definitions is that the most repelling LCS  definition involves  the eigenvectors and eigenvalues of the Cauchy-Green strain tensor, while the second condition in the definition of the second derivative ridge  is governed by the eigenvectors and eigenvalues of  the matrix $\Sigma$.
\section{Example: a Hamiltonian flow map}
\begin{figure}
\centerline{\includegraphics[width=0.7\columnwidth]{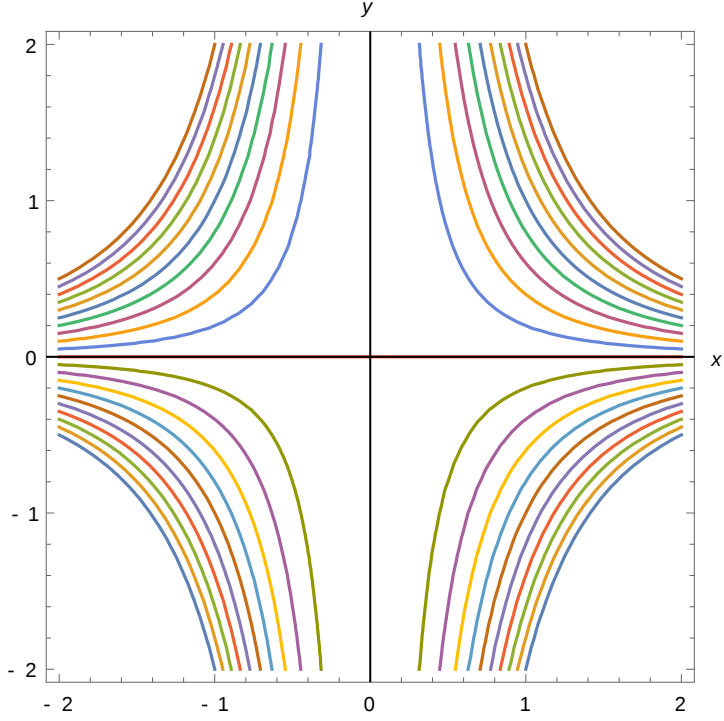}}
  \caption{Contour plot of the Hamiltonian  shown in Eq. (\ref{eq Hamiltonian}).}
  \label{fig: b}
\end{figure}

\begin{figure}
  \centerline{\includegraphics[width=0.7\columnwidth]{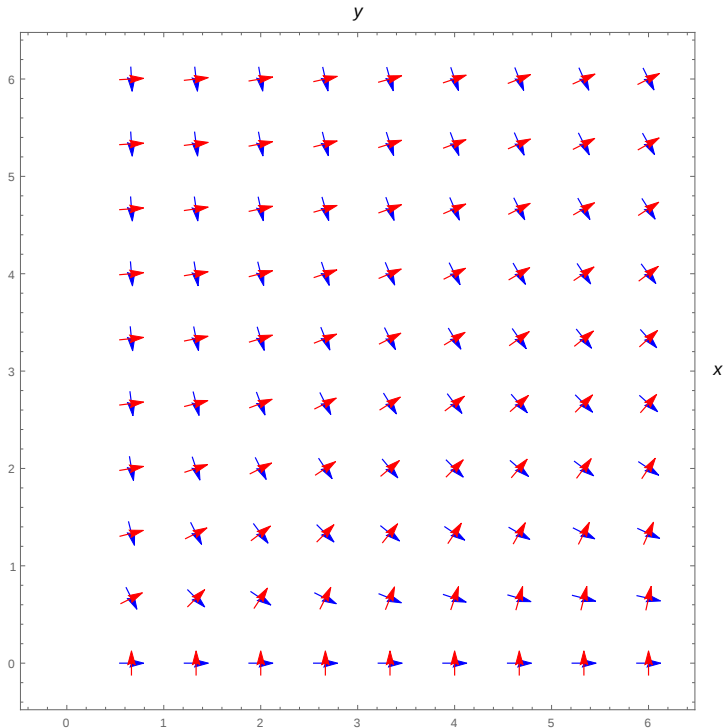}}
  \caption{Plot of the eigenvectors of the \emph{Cauchy-Green strain tensor} in the limit $t x_0 \gg 2$. $| \xi_{max} \rangle $ is marked in red while $| \xi_{min} \rangle $ in blue.}
  \label{fig: c}
\end{figure}

In this section we will illustrate the definitions introduced above  and the procedure needed in order to identify and characterize  the \emph{Lagrangian coherent structures} by direct construction in  a simplified dynamical system  that  can be studied analytically. We find it convenient to consider a Hamiltonian system which ensures condition (\ref{condizionelambda})   through conservation of phase space volume.
We consider the time independent  Hamiltonian 
\begin{equation}
  \label{eq Hamiltonian}
  H=\frac{x^2y}{2},
\end{equation}
from which we obtain 
\begin{equation}
  \label{eq esempio2}
  \left\{\begin{array}{ll}

      \frac{dx}{dt}= - \frac{x^2}{2}\\ \frac{dy}{dt}=xy

    \end{array}\right.
\end{equation}
A contour plot of the Hamiltonian $H$ is given in Fig. \ref{fig: b}. The trajectories obtained by integrating Eq. (\ref{eq esempio2}) coincide with the lines of constant $H$.
Integrating the system (\ref{eq esempio2}) we obtain
 \begin{eqnarray}
  \label{eq solutions3}
 x(t)  =  \left({\boldsymbol \phi}_0^t({ x}_0,y_0)\right)_x &=&\frac{2 { x}_0}{2 + { x}_0 t} \\
  y(t)  =  \left({\boldsymbol \phi}_0^t({  x}_0,y_0)\right)_y &=&y_0 \left( 1 + \frac{{  x}_0 t}{2}\right)^2,
\end{eqnarray}
where ${ x}_0 = x(t=0) $ and $y_0=y(t=0) $. 
From  Eq. (\ref{eq solutions3}) we see that points with negative ${  x}_0$ reach $x = - \infty$ in a finite time while points with positive  ${  x}_0$ reach $x = 0$  for $t \to \infty$. In order to avoid this finite time singularity  we restrict the domain to the semi-plane ${  x}_0 >0$. \\ The matrix $\nabla {\boldsymbol \phi}^t_0$ takes the form

\begin{equation}
  \label{eq jacobianmatrix2}
  \nabla {\boldsymbol \phi}_0^t =
  \left(\begin{array}{cc}
{1}/{B^2}\,  \, & 0\\ 
{B}ty_0  \,  \,& {B^2}
  \end{array}\right)
\end{equation}
where  $B=1+ t {  x}_0/2 $.
The \emph{Cauchy Green strain tensor} $C_0^t({  x}_0,y_0) \equiv \left( \nabla {\boldsymbol \phi}_0^t\right)^T  \, \nabla {\boldsymbol \phi}_0^t $ is
\begin{equation}
  \label{eq cauchygreen}
  C_0^t({  x}_0,y_0)=  B^2
  \left(\begin{array}{cc}
   {1}/{B^6} + \left( {ty_0}\right)^2 & \, \, \,  \,  B {ty_0}\\ 
B {ty_0}  & \, \, \, \,  B^2
\end{array}\right).
\end{equation}
In order to identify its eigenvectors and corresponding eigenvalues it may be convenient to diagonalize it  
\begin{equation}
  \label{eq Rotation}
R({\varphi})\, \, C^t_0 \, \, R^T({\varphi}) = D.
\end{equation}
by means of the  rotation  $R({ \varphi})$ 
\begin{equation}
  \label{eq rotat}
R({ \varphi}) =
\left(\begin{array}{cc}
 \cos{\varphi} & \, \, \,  \,  -\sin{\varphi}  \\ 
 \sin{\varphi}   & \, \, \, \,   \cos{\varphi} 
\end{array}\right)
\end{equation}
 with ${\varphi}$ given by 
\begin{equation}
  \label{eq phirotation}
  { \varphi}= \frac{1}{2} \arctan \left( \frac{2 B t y_0 }{{B}^2 -1/B^6- ( t y_0)^2}\right)
\end{equation} 
and $D ({  x}_0,y_0)$ is the  diagonal matrix
\begin{equation}
  \label{eq cauchydial }
  D =
  \left(\begin{array}{cc}
\lambda_{min} & \, \, \,  \, 0\\ 
0 & \, \, \, \, \lambda_{max} 
\end{array}\right)
\end{equation}
 with  $\lambda_{max} =  \lambda_{+}  $,  \, $\lambda_{min} =  \lambda_{-}  $\, and 
 \begin{equation}
  \label{eq eigenvaluessyst2}
  \lambda_{\pm}=\frac{1+ B^8 +   B^6 t^2 y_0^2  \pm 
\sqrt{\left[1+ B^8 +  B^6  t^2 y_0^2  \right]^2 - 4 B^8}}{2 \,B^4}.
\end{equation}
 The rotation angle $\varphi ({  x}_0, y_0) $ can be used to find the orientation of the strain lines.
Note that  $\varphi (x_0, y_0 = 0) = 0$, as  the strain tensor in Eq. (\ref{eq cauchygreen})  is diagonal  on the positive semi axis $y_0 = 0$  with $| \xi_{\min} \rangle $ tangent  this axis i.e., in the $x_0$ direction.
In fact from Eqs. (\ref{eq solutions3}) we see that if we take two points $x_{01}$ and $x_{02}$ on the positive  $y_0=0$  semi-axis, with $x_{02}>x_{01}$  they stay  on this axis  and their distance $d(t)$ decreases in time:
\begin{equation}
  \label{eq scaling}
  d(t)=\frac{d(0)}{1+ t\left(x_{02}+x_{01}\right)/2+ t^2 x_{02}x_{01}/4},
\end{equation}
as consistent with the fact that this $y_0  = 0$ axis  is parallel  to $| \xi_{\min} \rangle $ at all times.
In order to show the strain lines of the Cauchy-Green tensor for $x_0 > 0$  in Fig. \ref{fig: c} we plot the vector fields $| \xi_{max} \rangle  =  R^T({\varphi}) | 0,1 \rangle  =  | ( \sin{\varphi}, \cos {\varphi})  \rangle  $ and  $|\xi_{min} \rangle  =  R^T({\varphi}) |1,0 \rangle  = |\cos { \varphi}, -\sin {\varphi})  \rangle  $  in the limit of sufficiently long times such that $t x_0 \gg 2$  in which limit  the rotation  angle  $\varphi$ reduces to:
\begin{equation}  
\label{eq phirotation1} {\varphi} \sim (1/2)  \arctan[ ( x_0 y_0 )/(y_0^2 - x_0^2/4) ]= \arctan \frac{2 y_0}{x_0}
\end{equation} 
and becomes independent of time.
In the same limit the eigenvalues of the Cauchy-Green strain tensor become simply 
\begin{equation}
  \label{eq eigenvaluesaftersometime}
  \lambda_{max}\sim {t^4}\left(  x_0^4 + 4 y_0^2 {  x}_0^2\right)/16, \qquad    \lambda_{min} = 1/\lambda_{max}
\end{equation}
and have  a factorized time and space dependence.

\begin{figure}
 \centerline{\includegraphics[width=0.7\columnwidth]{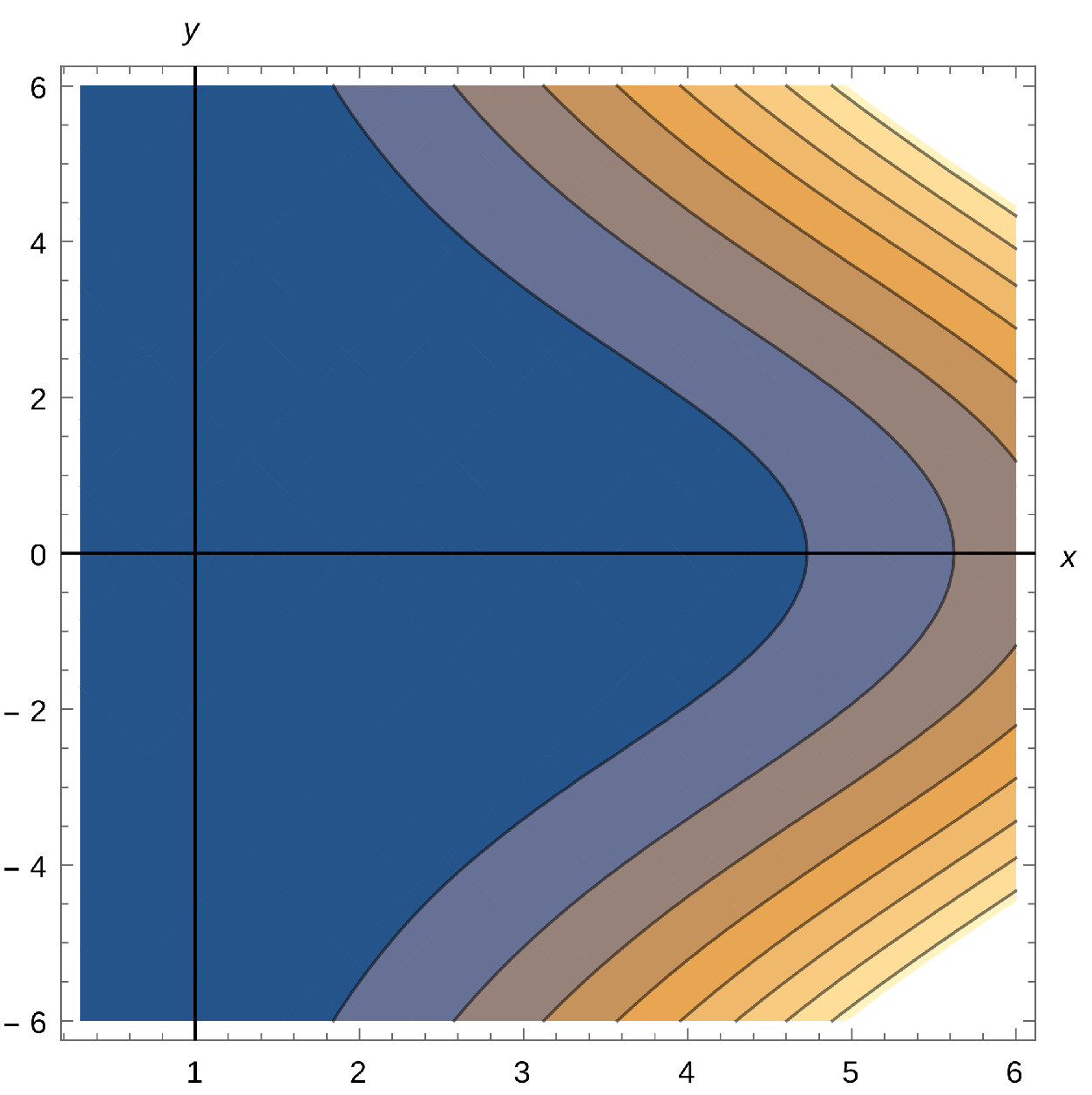}}
  \caption{Contour plot of  $\lambda_{max}.$}
  \label{fig: d}
\end{figure}

From (\ref{eq eigenvaluesaftersometime}) we can compute  ${\boldsymbol \nabla} \lambda_{max}$  (which will be parallel to ${\boldsymbol \nabla}  \lambda_{min}$)
\begin{equation}
  \label{eq nablalambda2}
{\boldsymbol \nabla}  \lambda_{max}= \frac{t^4 x_0}{4}
\left(\begin{array}{cc}
x_0^2 + 2 y_0^2\\ 
2x_0y_0
\end{array}\right) \quad \propto  \cos{\varphi} \left(\begin{array}{cc}
\cos^2{\varphi} + 2 \sin^2{\varphi} \\
2 \cos{\varphi} \sin{\varphi}
\end{array}\right) .\end{equation}
A contour plot of Eq. (\ref{eq eigenvaluesaftersometime}) is shown in Figure \ref{fig: d}. 
Recalling the sufficient and necessary condition in order to have a \emph{WLCS}  is $\left \langle \xi_{max} \Big |  \nabla \lambda_{max} \right \rangle=0$ we 
find the condition $\cos{\varphi} \,  \sin{\varphi}\, ( 2 \sin^2{\varphi} + 3 \cos^2{\varphi}) = 0$. 
For $x_0 \not= 0$ (i.e $\varphi \not= \pi /2$),  this condition   implies $\sin{\varphi} =0$.
Thus the only repulsive \emph{WLCS} is  given by the positive $y_0 = 0$ semi-axis. 

It is easy to see that the condition (\ref{maxrep}) is not satisfied, i.e., this   \emph{WLCS}
is not a maximum for the \emph{repulsive rate} and thus is not an LCS. 

Referring now to the definition of the second derivative ridges,  from \ref{eq eigenvaluesaftersometime} at $y_0 = 0$ we find 
\begin{equation}  \Sigma (x_0, y_0 = 0) =  \label{ridge1} \frac{4}{ x_0^2}
  \left(\begin{array}{cc}
-1 & \, \, \,  \, 0  \\ 
\, \, 0 & \, \, \, \,   2
\end{array}\right)
\end{equation}
which is not negative definite and for which $ \left \langle  n  \left|  \Sigma \right| n \right \rangle $  is a maximum, not a minimum, contrary to the  requirement (\ref{Sigma})  in the second derivative ridge definition.

In this example the WLCS according to Haller's definition that we found at $y_0 =0$ is not a second derivative ridge.
\section{Electron transport in a 3D reconnection process}

In this section we will deal with the barriers to the transport of electrons in a $3$D magnetic reconnection processes in the presence of a fixed ``guide field'' component $B_z$. 
A simplified doubly-periodic slab  geometry is considered. Because of this  double periodicity  this model can be applied mainly to toroidal thermonuclear plasmas \citep{borgogno2005aspects, avinash1998forced, porcelli2002recent}, and with  some modification  to space plasmas \citep{rappazzo2013current} and, more generally, to physical contexts where plasma is confined by  a nearly uniform  strong magnetic field with small perturbations. Under proper adiabaticity conditions, and in particular assuming that their  Larmor radius is negligible, charged particles move  along magnetic field lines. Therefore in this approximation  the study of the motion of electrons  can be referred to the study of the topology of the magnetic field lines. 

The  magnetic field structure and evolution  are described through the evolution of the magnetic flux function $\psi(x,y,z,t)$  by the relationship
\begin{equation}
{\bf B}= B_0{\bf e}_z + {\bf e}_z\times  \nabla\psi.
\end{equation} 
As for all solenoidal fields in an odd dimensional space, at any fixed time $t = {\bar t} $  the magnetic field line equations can be written in the form of Hamilton's equations.  The function  $\psi(x,y,z,{\bar t})$ plays the role of  the Hamiltonian while the coordinate $z$ plays the role of the time variable in this Hamiltonian system. Using a periodic  geometry along $z$ allows us to study the shape of magnetic field lines  with dynamical systems techniques such as the Poincar\'e map. The growth and interaction of different unstable modes will naturally lead to the formation of structures in the magnetic field such as magnetic islands and to   chaotic behaviour of field lines. 
Assuming, as commonly done, that the time it takes electrons  to complete a number of  turns along $z$ before  the magnetic field  is significantly changed,  we can investigate  the advection of particles in such a chaotic system at a fixed  $t = {\bar t}$ using the LCS technique. This requires that the electron thermal velocity be sufficiently larger than the Alfv\`en speed  and allows us to highlight the finite time transport barriers of the system. Choosing  the LCS characteristic time span $
\left| z-z_{0}\right|$  properly, we obtain the boundaries of the regions where fast electron mixing is expected. After a time span of the order of the Alfv\'en time  we can expect 
that the shape of the magnetic field lines will have  changed significantly and therefore we need to ``refresh'' the Hamiltonian $\psi(x,y,z,{\bar t})$  and  plot a new Poincar\'e map for the magnetic field. 
The LCS of the Poincar\'e map calculated with a sufficient number of iterations marks  the boundaries of the regions where the electrons will mix.

In this paper  we will not deal with the numerical integration of the PDE governing the dynamics. We will instead study the shape of the magnetic field lines at fixed $t$ extracting $\psi$ from the numerical simulation carried out by \citet{borgogno2005aspects}. An analogue analysis has been carried out by \citet{borgogno2011barriersa} where the LCS have been calculated using the definition based on the ridges of the Finite time Lyapunov exponent field. We choose the same values for the parameters defining the LCS in order to make a comparison with this work.

\subsection{The physical system}
The physical system studied is a Hamiltonian reconnection process in a dissipationless 3D plasma immersed in a strong, uniform, externally imposed magnetic field.  A slab geometry is used with a formal  additional  periodicity along $x$ imposed for numerical convenience. The algorithm applied in the numerical simulation is detailed in \citet{borgogno2008stable}. 
The reconnection process develops in a  static equilibrium configuration given by:
\begin{equation}
\psi_{eq}(x)=A\cos(x),
\end{equation}
with $A=0.19$. The integration domain is defined by $-L_x<x<L_x$, $-L_y<y<L_y$, $-L_z<z<L_z$ with $L_x=\pi$, $L_y=2\pi$ and $L_z=16\pi$. The equilibrium is perturbed as \begin{equation}
\Psi(x,y,z,t)=\psi_{eq}(x)+\psi(x,y,z,t),
\end{equation}
where $\psi$ is  written as a sum over Fourier modes:
\begin{equation}
\psi(x,y,z,t)= \Sigma_i \, \psi_i(x,k_{yi}y+k_{zi}z,t)
\end{equation}
with $k_{yi}=\frac{2 \pi m_i}{L_y}$, $k_{zi}=\frac{2 \pi n_i}{L_z}$. The  field line equations  can be cast in Hamiltonian form:
\begin{equation}
\left\{\begin{array}{ll}\frac{dx}{dz}=\frac{1}{B_0}\frac{\partial \Psi}{\partial y}= \frac{B_x}{B_0}\\\frac{dy}{dz}=-\frac{1}{B_0}\frac{\partial \Psi}{\partial x}= \frac{B_y}{B_0}\end{array}\right.
\end{equation}
Because of  its periodicity along $z$, the system  can be paired to its Poincar\'e map and therefore we will deal with a 2D discrete time system instead of a 3D continuous one. 

\begin{figure}
\centerline{\includegraphics[width=1\columnwidth]{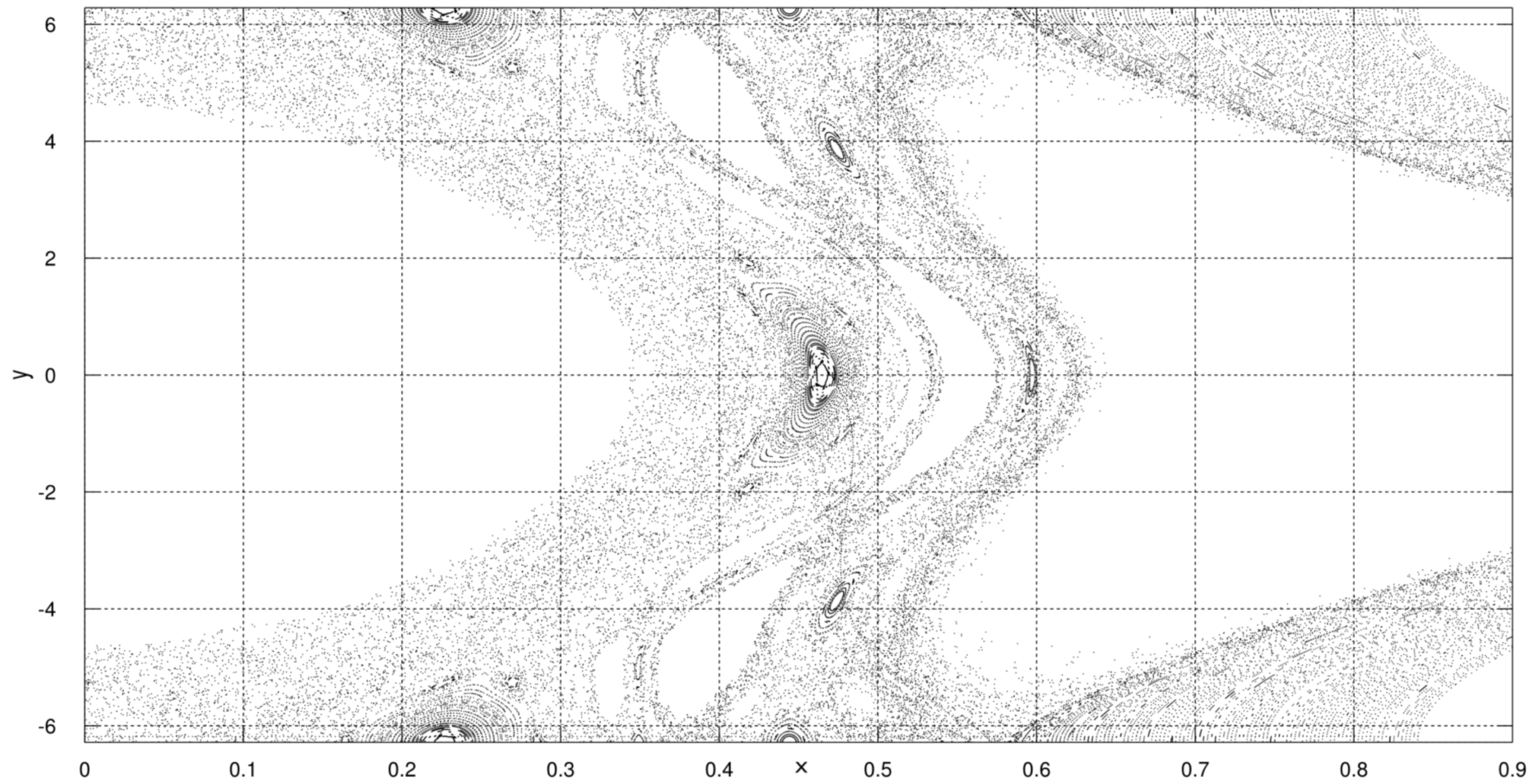}}
\caption{Poincar\'e  plot of the magnetic field lines obtained taking the intersections between the Magnetic field lines and the surface $z=L_{Z}/2$ with $t=415 \tau_A$.}
  \label{fig: e}
\end{figure}

The resonant condition  $\vec{B}_{eq}\cdot \vec{\nabla}\Psi=0$  leads to the definition of the resonant surfaces $x=x_{si}$ where the reconnection process takes place:
\begin{equation}
\frac{d\psi_{eq}(x)}{dx}=-\frac{\partial \psi_i/\partial z}{\partial \psi_i/\partial y}=-\frac{k_{zi}}{k_{yi}}.
\end{equation} 
The initial perturbation considered by \citet{borgogno2005aspects} consists of two contributions with different wave number pairs  $(k_{zi},\, k_{yi} )$:\,  $(1,0)$ and $(1,1)$  respectively
\begin{equation}
\psi(x,y,z,t)=\hat{\psi}_1(x,t)\exp(ik_{y1}y + ik_{z1}z)+\hat{\psi}_2(x,t)\exp(ik_{y2}y + ik_{z2}z).
\label{perturbations}
\end{equation}
The amplitude of the perturbation $\hat{\psi}_1$ is of order $10^{-4}$ while $\hat{\psi}_2$ is of order $10^{-5}$. Magnetic islands are induced around resonant surfaces $x=x_{si}$, i.e. $x_{s1}=0$ and $x_{s2}=0.71$. When magnetic islands are sufficiently large the islands  start interacting. Different modes of higher order are generated and the magnetic field topology becomes chaotic. The whole process can be visualized with the Poincar\'e  map technique which  at  any fixed  $t$ draws a snapshot of the magnetic field lines passing through a fixed $z$ plane. Following \citep{borgogno2005aspects} we consider the system after a time lapse of $t=415 \tau_A$ obtaining the Poincar\'e  map shown in Figure \ref{fig: e} with the intersections between the magnetic field lines and the $z=L_{z}/2$ surface. 
\subsection{Numerical results and comparison with FTLE ridges}
\begin{figure}
\centerline{\includegraphics[width=1\columnwidth]{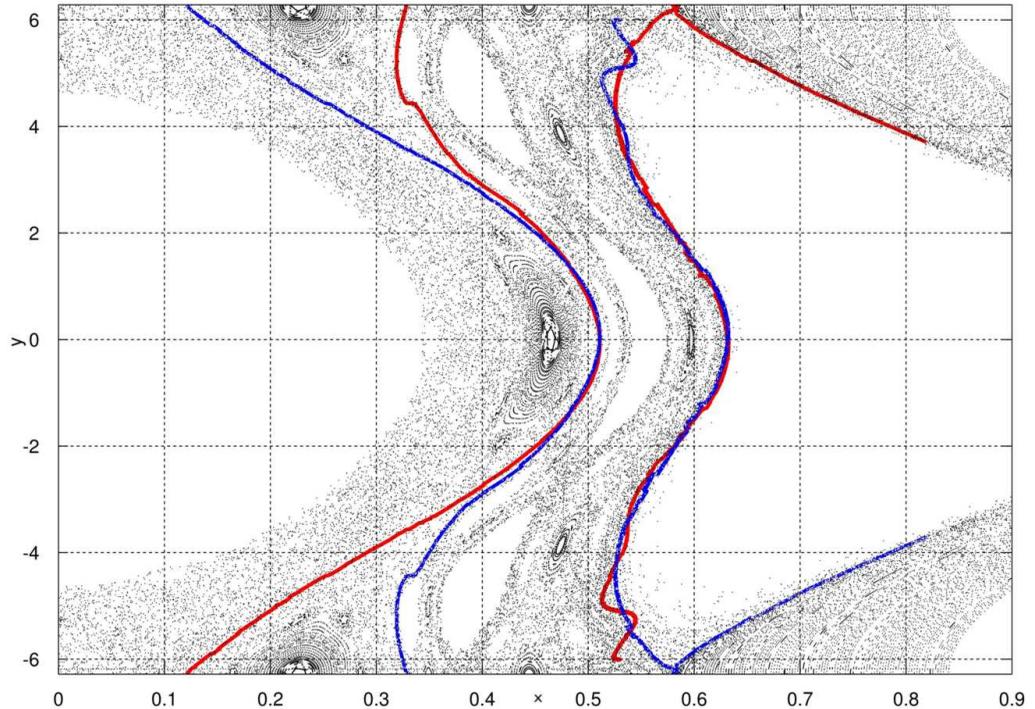}}
\caption{Plot of the most repelling material lines in red and of the most attractive ones in blue on the $z=L_z/2$ plane superimposed to the Poincar\'e map of the system. There is good agreement between the structures of the map and the material lines.}
\label{fig: f}
\end{figure}

\begin{figure}
\centerline{\includegraphics[width=1\columnwidth]{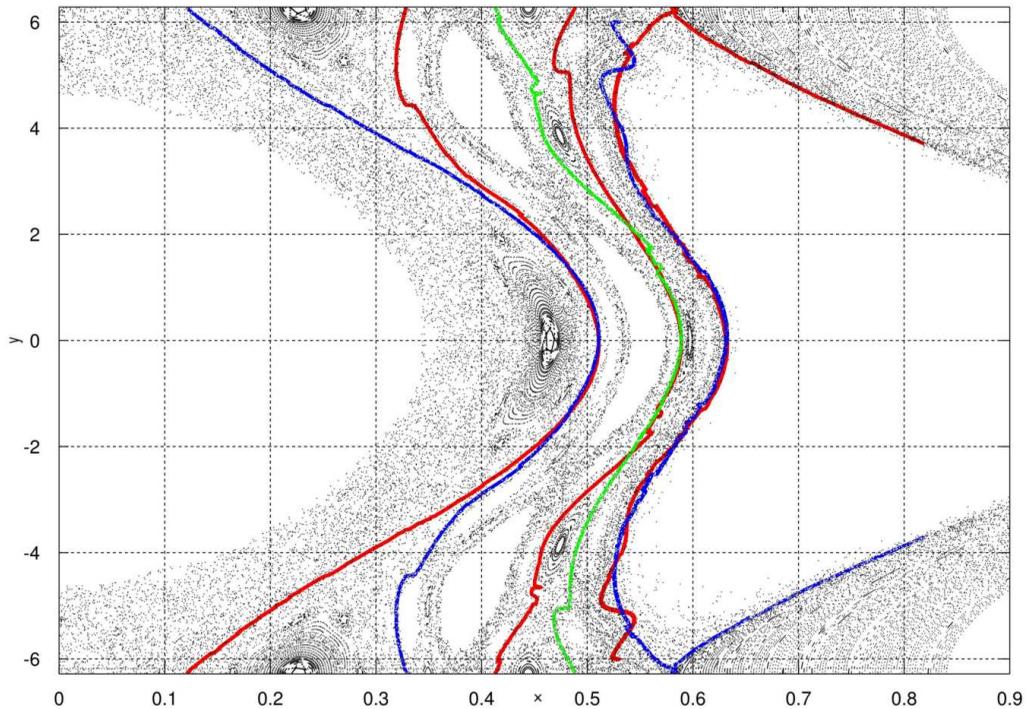}}
\caption{As in Figure \ref{fig: f} but with a lower filtering parameter. Repelling structures are marked in red while the attracting ones with other colors.}
\label{fig: g}
\end{figure}

We look for the LCS of the system defined as the most repelling material lines. The finite transport barriers of the system also include  the most repelling material lines obtained integrating backward in time, i.e. the attractive structures. We exploit  the symmetry of the system under reflection over $y=0$ to compute only the repulsive structures obtaining the attractive ones through their reflection. The algorithm used  to integrate  the dynamical system defined above is described by \citet{onu2015lcs}.  The result is shown in Figure \ref{fig: f} where red lines represent the repulsive structures while the blue ones the attractive ones.
The domain can be split into different regions which have the LCS as boundaries. We expect fast mixing processes inside these regions. The number of material lines drawn depends on the filtering parameter chosen as explained by \citet{onu2015lcs}. Lowering the filtering we obtain additional lines as depicted in Figure \ref{fig: g}. A comparison between the structures obtained and the FTLE ridges shown in \citep{borgogno2011barriersa,borgogno2011barriers} shows a  significant qualitative agreement.
\section{Conclusions}
This article is meant as an introduction to the study of Lagrangian Coherent Structures in chaotic magnetic field configurations. It aims to: \\
1) stress  the role  that Lagrangian Coherent Structures  can play in the description of transport phenomena in magnetically confined plasmas,\\
2) introduce the plasma physics reader to a debate, occurring mostly in the fluid dynamics and oceanographic communities, on the proper operational definitions of these structures,\\
3) exemplify the differences between the two definitions given by \citep{haller2011variational} and by \citep{shadden} on a simple, analytically solvable, case, \\
4) present  a preliminary numerical comparison between the results given in previous works \citep{borgogno2011barriersa, borgogno2011barriers} where the first definition given by\citet{shadden} was used and a recalculation of the same structures based on the corrected definition given by \citet{haller2011variational}. 

These results appear to be promising enough to start implementing a wider investigation of the applicability of LCS to magnetic configurations without the constraint of periodicity which is inherent to the Poincar\'e map method.
\bibliographystyle{jpp}

\bibliography{transport}
\end{document}